\documentclass[aps,preprint]{revtex4}%
\usepackage{amsfonts}
\usepackage{amsmath}
\usepackage{amssymb}
\usepackage{graphicx}%
\begin{document}
\title[Short title for running header]{The Driving Force of Superconducting Transition in High Temperature Superconductors}
\author{Tao Li}
\affiliation{Center for Advanced Study, Tsinghua University, Beijing 100084, P.R.China}

\begin{abstract}

We show that both the kinetic energy and the exchange energy of the t-J model
can be read off from the optical data. We show that the optical data indicates
that the superconducting transition in high temperature superconductors is
kinetic energy driven and the exchange energy resist the transition. We also
show that kinetic energy may also be the driving force of the pseudogap phenomenon.

\end{abstract}
\volumeyear{year}
\volumenumber{number}
\issuenumber{number}
\eid{identifier}
\startpage{1}
\endpage{10}
\maketitle

\bigskip The question of the driving force of the superconducting transition
in high temperature superconductors is hotly debated upon in recent
years\cite{1,2,3,4,5,6,7}. Two school of thoughts compete. In terms of the t-J
model, they are called kinetic energy driving mechanism and exchange energy
driving mechanism. Since the competition of the kinetic energy and the
exchange energy is the key to understanding the high-Tc physics in the t-J
model perspective, an unambiguous answer to the driving force question is important.

In conventional BCS superconductors, the superconducting transition is driven
by some attractive pairing potential between electrons and the kinetic energy
is frustrated in the paired superconducting state. In the high temperature
superconductors, the kinetic energy is frustrated in the half-filled parent
compounds which are Mott insulators. Carrier doping releases the frustrated
kinetic energy and lead to high temperature superconductivity for sufficiently
large doping concentration. Hence a kinetic energy driving mechanism seems
quite reasonable at least on the underdoped side of the phase
diagram\cite{1,2,3,4}. However, arguments for exchange energy driving
mechanism can also be made since it is the antiferromagnetic exchange that
pair the electrons into local spin singlet. Especially, the SO(5) theory
claims that the superconducting transition in the high Tc superconductors is
driven by the exchange energy saving as the so called $\pi$ - resonance open a
new channel for antiferromagnetic spin fluctuation in the superconducting
state\cite{5,6,7}.

Experimentally, the kinetic energy can be measured with the help of the
optical sum rule. According to this rule, the absolute value of the kinetic
energy of a single band model\ is equal to the total intraband optical
transition rate, i.e.%

\[
\left\langle -K\right\rangle =%
{\displaystyle\int\nolimits_{0}^{\Omega}}
\sigma_{1}\left(  \omega\right)  d\omega
\]
here $K$ is the kinetic energy of the model, $\sigma_{1}\left(  \omega\right)
$ is the real part of the optical conductivity, $\Omega$ is the energy cutoff
for intraband transition. In the case of the t-J model, an intraband
transition is a transition within the subspace of no double occupancy. The
corresponding cutoff energy $\Omega$ is set by the gap between this subspace
and the subspace with doubly occupied sites, i.e. the Hubbard gap. Recently,
optical measurement on $BSCCO_{2212}$ find a spectral weight transfer from the
high energy part of the spectrum($10^{4}cm^{-1}$ to $2\times10^{4}cm^{-1}$) to
the low energy part of the spectrum(below $10^{4}cm^{-1}$ ) with decreasing
temperature in both optimally doped and underdoped samples\cite{4}. If we take
$10^{4}cm^{-1}$ as the energy cutoff of the intraband transition in the t-J
model, then the experimental result indicates that the kinetic energy is
lowered with decreasing temperature. This result is taken as evidence in
support of the kinetic energy driving mechanism. The decrease of the kinetic
energy in the superconducting state estimated from the optical data is about
$1meV$ per cooper atom, an order of magnitude larger than the superconducting
condensation energy estimated from thermodynamical measurements\cite{8}.

However, the neutron scattering experiments tell a quite different
story\cite{7}. The neutron scattering experiments measure the spin fluctuation
spectrum of the system. By integrating the spectrum with suitable weighting
factor, the exchange energy can be deduced\cite{5}. This is expressed as the
following sum rule%

\[
\left\langle S_{i}\cdot S_{j}\right\rangle =3%
{\displaystyle\int}
\frac{dq^{2}}{\left(  2\pi\right)  ^{2}}%
{\displaystyle\int\nolimits_{0}^{\infty}}
\frac{d\omega}{\pi}\operatorname{Im}\chi\left(  q,\omega\right)  \cos
[q\cdot(i-j)]
\]
where $\operatorname{Im}\chi\left(  q,\omega\right)  $ is the spin fluctuation
spectrum of the system. In high temperature superconductors, the spin
fluctuation spectrum develops a resonant mode around $\left(  \pi,\pi\right)
$ in the superconducting state\cite{9}. This mode give a negative contribution
to the exchange energy which is argued by some author to be the source
of the superconducting condensation energy\cite{6,7}. The superconducting
condensation energy estimated in this way is also an order of magnitude larger
than that estimated from thermodynamical measurements. Obviously, the
conclusions reached by optical measurements and the neutron measurements can
not be both correct. Since optical data has higher resolution and covers
larger energy range as compared with the neutron data, we expect the optical
conclusion to have a better chance to survive.

The key observation made in this paper is that both the kinetic energy and the
exchange energy in the t-J model are kinetic energy in nature and can both be
measured with optics. This is almost obvious if we realize that the t-J model
is in fact a low energy effective theory of an underlying Hubbard-type model.
The kinetic energy term in the t-J model corresponds to the real kinetic
process within the subspace of no double occupancy while the exchange energy
term corresponds to the virtual kinetic process which involve doubly occupied
sites in intermediate step. Here we take the standard Hubbard model to
demonstrate the idea%

\[
H=H_{t}+H_{U}=-t%
{\displaystyle\sum\limits_{\left\langle i,j\right\rangle ,\sigma}}
(c_{i,\sigma}^{\dagger}c_{j,\sigma}+h.c.)+U%
{\displaystyle\sum\limits_{i}}
n_{i,\uparrow}n_{i,\downarrow}%
\]
In the large $U$ limit, the Hubbard model reduce to the t-J model in the low
energy subspace of no double occupancy. This is usually done by dividing the
Hilbert space into the low energy subspace of no double occupancy and the high
energy subspace with nonzero doubly occupied sites. In the large $U$ limit,
the two subspaces are separated by a gap of order $U$. Correspondingly, the
Hamiltonian can be divided into intra-subspace pieces and inter-subspace
pieces. Introducing the projection operator $P_{L}$ and $P_{H\text{ }}$for the
low energy subspace and the high energy subspace, the Hubbard Hamiltonian can
be written as%

\[
H=H_{L}+H_{H}+H_{mix}%
\]
in which%

\begin{align*}
H_{L} &  =P_{L}H_{t}P_{L}\\
H_{H} &  =P_{H}H_{t}P_{H}+H_{U}\\
H_{mix} &  =P_{L}H_{t}P_{H}+P_{H}H_{t}P_{L}\\
&
\end{align*}
are the Hamiltonian in the low energy subspace, Hamiltonian in the high energy
subspace and the subspace-mixing term. The subspace-mixing term can be removed
by a canonical transformation $e^{iS}$. To first order of $\frac{t}{U}$, the
transformed Hamiltonian in the low energy subspace is the standard t-J model%

\begin{align*}
e^{iS}He^{-iS} &  =H_{t-J}=P_{L}H_{t}P_{L}+\frac{i}{2}\left[  S,H_{mix}%
\right]  \\
&  =-t%
{\displaystyle\sum\limits_{\left\langle i,j\right\rangle ,\sigma}}
(\hat{c}_{i,\sigma}^{\dagger}\hat{c}_{j,\sigma}+h.c.)+J%
{\displaystyle\sum\limits_{\left\langle i,j\right\rangle }}
\left(  S_{i}\cdot S_{j}-\frac{1}{4}n_{i}n_{j}\right)
\end{align*}
where $\hat{c}_{i,\sigma}=\left(  1-n_{i,\bar{\sigma}}\right)  c_{i,\sigma}$ ,
$\vec{S}_{i}=\frac{1}{2}c_{i,\alpha}^{\dagger}\vec{\sigma}c_{i,\beta}$,
$J=\frac{4t^{2}}{U}$. Under the canonical transformation, the kinetic energy
of the Hubbard model transforms into%
\[
e^{iS}H_{t}e^{-iS}=-t%
{\displaystyle\sum\limits_{\left\langle i,j\right\rangle ,\sigma}}
(\hat{c}_{i,\sigma}^{\dagger}\hat{c}_{j,\sigma}+h.c.)+2J%
{\displaystyle\sum\limits_{\left\langle i,j\right\rangle }}
\left(  S_{i}\cdot S_{j}-\frac{1}{4}n_{i}n_{j}\right)
\]
in the low energy subspace to first order of $\frac{t}{U}$. The potential
energy of the Hubbard model transforms into %

\[
e^{iS}H_{U}e^{-iS}=-J%
{\displaystyle\sum\limits_{\left\langle i,j\right\rangle }}
\left(  S_{i}\cdot S_{j}-\frac{1}{4}n_{i}n_{j}\right)
\]
\ in the low energy subspace to the same order.\ Note the transformed form of
$H_{U}$ is still positive definite. From these formulas, we see explicitly the
relation between the charge response in the t-J model and that of the
underlying Hubbard model. Especially, we see how the exchange term of the t-J
model(which is charge neutral in the subspace of no double occupancy)
contribute to the charge response of the underlying Hubbard model. According
to the optical sum rule, the kinetic energy of the Hubbard model is related to
its total optical response in the following way%

\[
\left\langle -e^{iS}H_{t}e^{-iS}\right\rangle =%
{\displaystyle\int\nolimits_{0}^{\Lambda}}
\sigma_{1}\left(  \omega\right)  d\omega
\]
where $\Lambda$ is the energy cutoff for the Hubbard model. Here we assume
that other bands of the system are far away from the Fermi surface and the
optical weight measured in experiment is due to a single band. The total
optical weight of the Hubbard model is composed of two contributions, namely
the optical response within the subspace of no double occupancy and the
optical response involving doubly occupied sites. The first contribution is
just the optical response of the t-J model. According to the optical sum rule,
this contribution is related to kinetic energy of the t-J model in the
following way%

\[
\left\langle t%
{\displaystyle\sum\limits_{\left\langle i,j\right\rangle ,\sigma}}
(\hat{c}_{i,\sigma}^{\dagger}\hat{c}_{j,\sigma}+h.c.)\right\rangle =%
{\displaystyle\int\nolimits_{0}^{\Omega}}
\sigma_{1}\left(  \omega\right)  d\omega
\]
. Hence the high energy optical weight, i.e. the optical weight between
$\Omega$ and $\Lambda$ is related to the exchange energy of the t-J model in
the following way%

\[
\left\langle -2J%
{\displaystyle\sum\limits_{\left\langle i,j\right\rangle }}
\left(  S_{i}\cdot S_{j}-\frac{1}{4}n_{i}n_{j}\right)  \right\rangle =%
{\displaystyle\int\nolimits_{\Omega}^{\Lambda}}
\sigma_{1}\left(  \omega\right)  d\omega
\]
Thus, if we can determine the energy cutoff $\Lambda$ and $\Omega$, we can
read off both the kinetic energy and the exchange energy of the t-J model from the
optical data. 

The energy cutoff $\Omega$ of the t-J model for cuprates is about
$10^{4}cm^{-1}$ where a conductivity minimum separate the intraband transition
and the interband transition\cite{4}. The value of the energy cutoff $\Lambda$
of the Hubbard model is subjected to some uncertainty. Here we adopt the value
$2\times10^{4}cm^{-1}$ below which reliable optical data is available\cite{4}.
Experimentally, the spectral weight between $10^{4}cm^{-1}$ to $2\times
10^{4}cm^{-1}$ transfers steadily to below $10^{4}cm^{-1}$ with decreasing
temperature with the total spectral weight approximately conserved\cite{4}.
According to our analysis in the last paragraph, this indicates that the
kinetic energy of the t-J model decreases steadily with decreasing temperature
while the exchange energy increases steadily with decreasing temperature.
Closer examination on the data shows that both energies change more rapidly
upon entering the superconducting state. This shows convincingly that the
superconducting transition is kinetic energy driven while the exchange energy
resist the transition. Also, we note since the kinetic energy and the exchange
energy evolve in opposite direction with temperature, the true superconducting
condensation energy should be smaller than that estimated from the kinetic
energy alone.

Finally, we note the optical measurement can also serve as a probe of driving
force for other phenomena in high temperature superconductors such as
pseudogap, stripes and \textit{et al.} For example, the steady increase of the
exchange energy with decreasing temperature seems at odd with the conventional
understanding of the pseudogap as some kind of spin pairing gap driven the
exchange energy of the t-J model\cite{10}. This indicates that the kinetic
energy of the t-J model plays an important role in the formation of the
pseudogap\cite{11,12}. In this respect it is interesting to do optical
measurement on overdoped samples where the spin gap and the superconducting
transition are believed to be exchange energy driven.

\bigskip The author would like to thank members of the HTS group at CASTU for discussion.

\end{document}